%% file: ms.tex
\documentclass[letterpaper, 10 pt, conference]{ieeeconf}

\IEEEoverridecommandlockouts                             
\usepackage{cite}

\usepackage[pdftex]{graphicx}
\usepackage{subcaption}
\usepackage{multirow}
\usepackage{pgfplots}
\pgfplotsset{width=7cm,compat=newest}
\usepackage{textcomp}

\usepackage{amsmath}
\usepackage{units}
\usepackage[nolist]{acronym}
\usepackage{array}

\newcommand{\disable}[1]{}

\input{corporateColours.tex}

\begin{document}

\title{\LARGE \bf
Benchmarking Image Sensors Under Adverse Weather Conditions for 
Autonomous Driving}

\author{Mario Bijelic$^{*}$, Tobias Gruber$^{*}$ and Werner Ritter%
\thanks{All authors are with Daimler AG, Wilhelm-Runge-Str. 11, 89081 Ulm, Germany. \{mario.bijelic, tobias.gruber, werner.r.ritter\}@daimler.com}
\thanks{$^{*}$ Mario Bijelic and Tobias Gruber have contributed equally to the work.}
}

\maketitle
\thispagestyle{empty}
\pagestyle{empty}

\begin{acronym}
 \acro{CMOS}{complementary metal-oxide semiconductor}
 \acro{EU}{European Union}
 \acro{DENSE}{aDverse wEather eNvironment Sensing systEm}
 \acro{FIR}{far infrared}
 \acro{NIR}{near infrared}
 \acro{SWIR}{short wave infrared}
 \acro{ADAS}{automotive drive assistance system}
 \acro{RMS}{root mean squared}
 \acro{ZNCC}{zero-mean normalized cross correlation}
 \acro{HDR}{high dynamic range}
\end{acronym}

\input{abstract}
\input{introduction}
\input{theory}
\input{evaluated_sensors}
\input{fogchamber_experiment}
\input{evaluation}
\input{outlook_conclusion}
\input{acknowledgment}

\bibliographystyle{IEEEtran}
\bibliography{references}

\end{document}

%% file: corporateColours.tex
\definecolor{dai_ligth_grey}{RGB}{230,230,230}
\definecolor{dai_ligth_grey20K}{RGB}{200,200,200}
\definecolor{dai_ligth_grey40K}{RGB}{158,158,158}
\definecolor{dai_ligth_grey60K}{RGB}{112,112,112}
\definecolor{dai_ligth_grey80K}{RGB}{68,68,68}

\definecolor{dai_petrol}{RGB}{0,103,127}
\definecolor{dai_petrol20K}{RGB}{0,86,106}
\definecolor{dai_petrol40K}{RGB}{0,67,85}
\definecolor{dai_petrol80}{RGB}{0,122,147}
\definecolor{dai_petrol60}{RGB}{80,151,171}
\definecolor{dai_petrol40}{RGB}{121,174,191}
\definecolor{dai_petrol20}{RGB}{166,202,216}

\definecolor{dai_deepred}{RGB}{113,24,12}
\definecolor{dai_deepred20K}{RGB}{90,19,10}
\definecolor{dai_deepred40K}{RGB}{68,14,7}


%% file: abstract.tex
\begin{abstract}
Adverse weather conditions are very challenging for autonomous driving because most of the state-of-the-art sensors stop working reliably under these conditions. 
In order to develop robust sensors and algorithms, tests with current sensors in defined weather conditions are crucial for determining the impact of bad weather for each sensor. 
This work describes a testing and evaluation methodology that helps to benchmark novel sensor technologies and compare them to state-of-the-art sensors.
As an example, gated imaging is compared to standard imaging under foggy conditions. 
It is shown that gated imaging outperforms state-of-the-art standard passive imaging due to time-synchronized active illumination.
\end{abstract}

%% file: introduction.tex
\section{Introduction}

In recent years, autonomous driving has experienced a tremendous buildup and sparked a fierce race between car manufacturers and computer companies. 
For safe autonomous driving, environment perception plays a key role: detection of pedestrians and cyclists, traffic lights and unexpected obstacles on the road are inevitable. 
Many different sensors such as cameras, radar, lidar and ultrasonic sensors can be used to perceive the environment as reliably as possible in order to recognize critical situations.

However, sensors such as stereo cameras often malfunction or even fail in adverse weather, when the accident risks are particularly high. 
The \ac{EU} project \ac{DENSE}\footnote{dense247.eu} focuses on developing and demonstrating an all-weather sensor suite for traffic services, driver assistance and autonomous driving. 

\begin{figure}[!t]
	\captionsetup[subfigure]{justification=centering}
	\centering 
	\begin{subfigure}[t]{\columnwidth}
		\centering
		\includegraphics[width=\columnwidth]{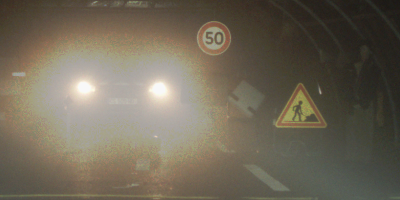}
		\caption{Standard \ac{CMOS} camera.}
	\end{subfigure}
	
	\begin{subfigure}[t]{\columnwidth}
		\centering
		\includegraphics[width=\columnwidth]{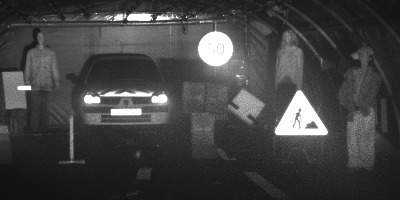}
		\caption{Gated camera.}
	\end{subfigure}
	
	\caption{Raw images of a standard \ac{CMOS} camera and a gated 
		camera taken in artificial fog in a fog chamber.}
	\label{fig:qualitative_comparison}
	
\end{figure}

There exist a bunch of emerging sensors in the automotive sector that are based on new technologies and wavelengths in order to tackle the problem of perception in bad weather situations. 
In order to investigate the advantages and disadvantages of these new sensors in all detail, testing methodologies and evaluation methods are required. 
Qualitative evaluation of visual sensors as in Fig.~\ref{fig:qualitative_comparison} is often quite easy but it often contains a lot of subjective perception and it is hard to quantize the advantages and disadvantages. 

This work presents a testing and evaluation methodology and describes how to benchmark different sensors in adverse weather conditions. As an example, we compare a standard \ac{CMOS} camera with a novel gated camera in the \ac{NIR}. All tests are performed under well-defined weather conditions in a fog chamber.

\input{related_work}

%% file: related_work.tex
\subsection{Related Work}  \label{sec:related_work}

Under adverse weather conditions, light is scattered by particles in the atmosphere which results in a decay of contrast \cite{Oakley1998,Schechner2001,Narasimhan2001}. 
There have been a lot of work on how to improve such degraded images. 
Oakley and Satherley reduced contrast degeneration by physical models if the scene geometry is known \cite{Oakley1998}. 
Schechner et al. used polarization in order to dehaze images \cite{Schechner2001} while Narasimhan and Nayar presented a simple method to restore the contrast by calculating the normalized radiance with a priori depth segmentation or two images under different weather conditions \cite{Narasimhan2001}. 
However, it is often quite subjective how good these algorithms improve the degenerated image.

New sensor technologies may also overcome this decay of contrast.
Nevertheless, these novel technologies have to be benchmarked against state-of-the-art technologies. 
Bernard et al. \cite{Bernard2013} developed a Monte-Carlo approach based on physical models in order to predict the performance of different imaging sensors under adverse weather conditions. 
Validating such a model with measurements in the natural environment is quite complicated as the fog or rain conditions are inhomogeneous and unstable. 
In recent years, climate chambers, e.g. CEREMA \cite{Colomb2008}, have emerged which are able to produce fog and rain in very stable conditions. 

In \cite{Duthon2016}, two evaluation methods for validating a rain simulator with real rain data in such a climate chamber are presented. 
In order to confirm the authenticity of the rain simulator, the average of standard deviations of small image patches and the \ac{ZNCC} between pairs of random image patches are used. 
However, these metrics rarely show a correlation to the rain intensity and the evaluation in foggy conditions is missing.

Hasirlioglu et al. describe in \cite{Hasirlioglu2016} a test methodology for rain influence on automotive sensors. They simulated different rain intensities by increasing the number of active rain layers and evaluated the performance of radar, lidar and camera. 

This work focuses on testing and evaluating novel automotive sensor technologies in fog. The methodology can be easily transformed to other adverse weather such as rain and snow or even to benchmark image enhancement algorithms.

%% file: theory.tex
\section{Theory} \label{sec:theory}

The theory of optics is essential to characterize light and its behavior in different weather conditions. 
The following briefly explains the main ideas and concepts in order to introduce the terminology which is needed for qualitative description. 
Moreover, a scattering model is derived which helps to validate the measurements later.

Different weather conditions are defined by different particles that are involved in a light scattering processes, e.g. small water droplets in fog and 
dirt particles in haze. 
These particles have certain physical properties, i.e. \emph{attenuation coefficient}, \emph{size} and \emph{thermodynamics}, which determine the disturbance. 
The size of the particles influences the scattering of light. 
\emph{Mie scattering} \cite{Wriedt2012} can be applied in any case as it is an exact solution if an incoming planar wave is scattered on a spherical obstacle. There exist approximations for special cases. 
If the light wavelength $\lambda$ is much smaller than the 
particle size $r$, i.e. $\lambda \ll r$, classical scattering is 
sufficient to describe the scattering process,
If $\lambda \gg r$, \emph{Rayleigh scattering} can be applied.
The effects of scattering can be observed each day in nature. 
Rayleigh scattering is responsible for the blue colored sky at daytime and the 
red sunsets \cite{YoungReighlexScattering}. 
Classical scattering cause \emph{coronas} or the \emph{Tyndall effect} and disturbs the imager on a higher level. 
A corona occurs in front of a light source. 
Assuming that the light source has an active area that emits light, the active area along a cone is increased by scattered light. 
The Tyndall effect is important when an indirect beam of light is scattered through fog or haze and the beam of light itself seems to be visible for the human observer. 
This effect can be observed in the forest when the sunbeams break through the tree crown and appear to be visible. 
In meteorology, all these effects are known as \emph{air-light} \cite{airlight2}. 

Attenuation which involves scattering and luminescence \cite{AttenuationDefinition} describes the loss of emitted light for the observer. 
For smaller and simpler scenarios, exact \emph{Monte Carlo} simulations can be used \cite{SemiMonteCarloDumont,CeremaFogCarSimulationDumont,RoadBeaconIlluminationFog}. 
However, such a method is very extensive. 
For the sake of simplicity, compared to \cite{SemiMonteCarloDumont,CeremaFogCarSimulationDumont} an easily observed model with an exponential scattering model is applied\cite{Narasimhan2002,OpenGlTut}. 

The intensity $I(d)$ on the chip can be explained by three main effects: scene radiance, attenuation and air-light. 
Scene radiance originates from the expansion of a point light source. 
The light intensity is distributed on a spherical surface and in this way the initial intensity $I_0$ decreases by a factor of $\frac{1}{d^2}$. 
Therefore, the scene radiance $J(d)$ can be described by
\begin{align}
 J(d)=\frac{I_0}{d^2}. \label{eq:scene_radiance}
\end{align}
Attenuation can be modeled by an exponential decay which depends on the attenuation coefficient $\beta$ and distance $d$. 
The intensity $I_\text{att}(d)$ at a certain distance $d$ is given by
\begin{align}
 I_\text{att}(d)=J(d)e^{-\beta d} = \frac{I_0}{d^2} e^{-\beta d}.\label{eq:attenuation}
\end{align}
As $I(d)$ decreases exponentially, the image gets darker and loses contrast in greater distance. 
The effect of air-light $A(d)$ is given by \cite{BookOpticsAthmosphere}
\begin{align}
I_\text{air}(d)=I_{\infty}\left(1-e^{-\beta d}\right),\label{eq:airlight}
\end{align}
where $I_{\infty}$ describes the horizon brightness. 
Finally, the scattering model can be obtained by adding both effects of attenuation and air-light, i.e.
\begin{align}
 I(d) = \frac{I_0}{d^2} e^{-\beta d} + 
I_{\infty}\left(1-e^{-\beta d}\right).\label{eq:scattering_model}
\end{align}
The air-light model forces the intensity to converge to $I_{\infty}$.

Colomb et al. introduce in \cite{Colomb2008} the visibility $V$, also called \emph{meteorological visual range}, as a practical parameter that takes into account the physical aspects of fog, the photometric aspects of objects, and visual perception.
It is derived from Koeschmieder's law \cite{Hinds1999}, namely
\begin{align}
V = -\frac{\log \epsilon}{\beta},\label{eq:visibility}
\end{align}
where $\epsilon$ is the contrast threshold, often assumed to be 5\,\%.

%% file: evaluated_sensors.tex
\section{Evaluated Sensors}

A test vehicle (CAR1) is equipped with the following sensors for a qualitative and quantitative comparison of sensor performance in bad weather situations:
\begin{itemize}
	\item \ac{CMOS} camera (Aptina AR0230)
	\item \ac{NIR} gated camera (BrightwayVision)
	\item \ac{NIR} lidar (Velodyne HDL-64 S2)
\end{itemize}

As a standard camera system, we use a \ac{CMOS} digital image sensor with resolution 1980x1088 pixels. It allows \ac{HDR} imaging with a frame rate of \unit[20]{Hz} and bit depth of \unit[12]{bit}.

In order to improve vision in adverse weather, gated imaging is a promising technology which originated from the military. 
The gated principle is based upon a time-synchronized camera and a pulsed laser \cite{Inbar2008}. 
The laser illuminates the scene for a very short time $t_\text{laser}$ and after a known delay $t_\text{delay}$ the imager is exposed for $t_\text{gate}$, which is called \emph{micro exposure}. 
On account of the speed of light, only photons from a certain distance in the scene can be captured on the imager. 
In order to obtain a sufficiently illuminated image, multiple micro exposures coming from a specific laser pulse are accumulated on the chip. 
After that, a \emph{slice}, which is a set of $m$ micro exposures, is read out. Gated imaging \cite{Inbar2008} has two main advantages:
\begin{enumerate}
 \item Backscattered light from clouds, fog or raindrops and from reflections on the ground are avoided due to the very short and time-synchronized integration time
 \item Additional depth information is obtained
\end{enumerate}
Since active illumination is required for this technology, many challenges have to be solved, e.g. eye-safety and interference with similar systems.
However, gated imaging is considered as a very promising technique for the automotive sector as it enables capturing undisturbed images even under adverse weather conditions where backscattering and reflections are common.

\begin{figure}
 \centering
 \includegraphics[width=\columnwidth]{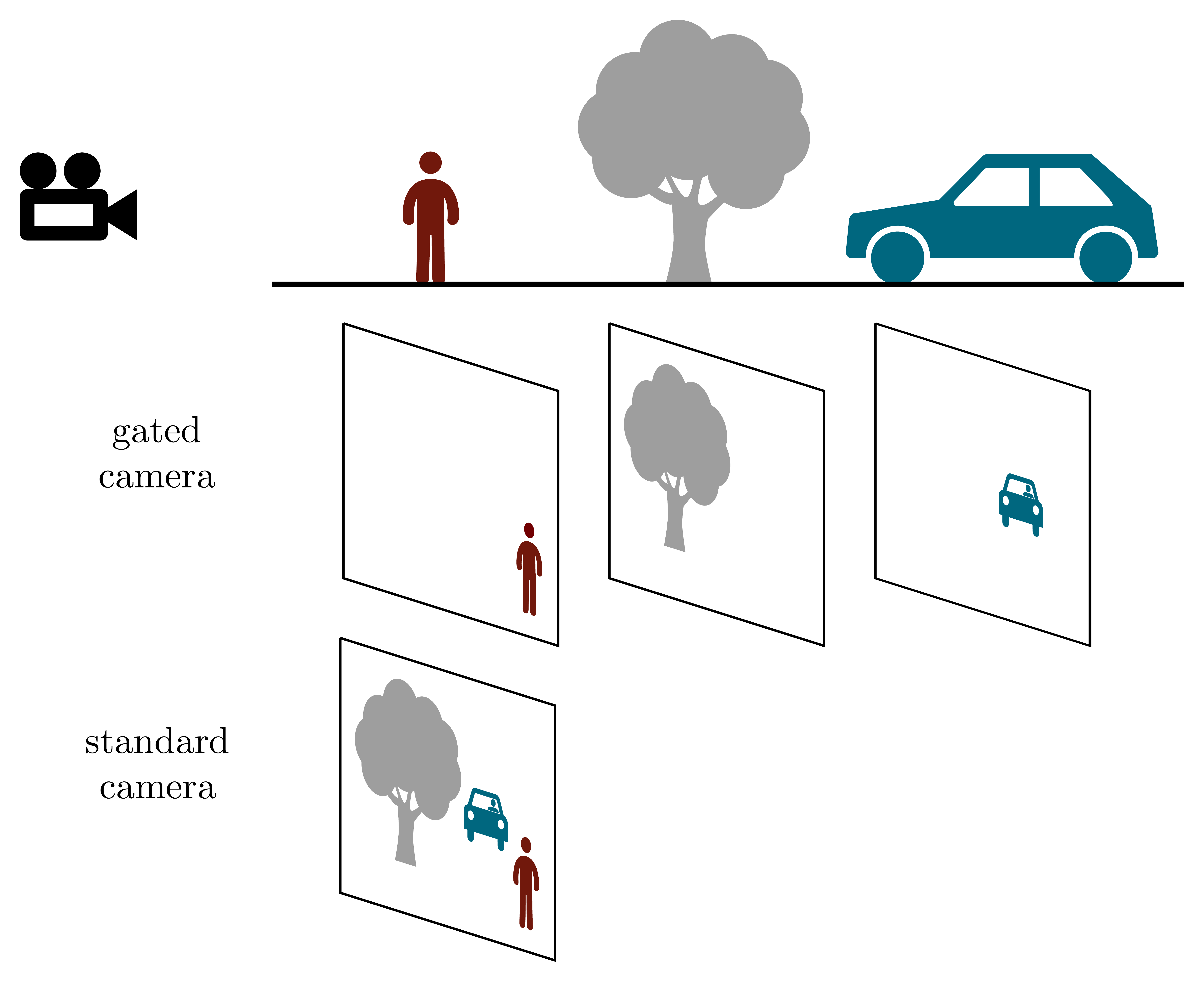}
 \caption{Illustration of slices taken from a full image.}
 \label{fig:gated_principle}
\end{figure}

Changing the parameters of the gating ($t_\text{laser}$, $t_\text{delay}$ and $t_\text{gate}$) paves the way to more interesting features. 
Suppose that there are three different gating schemes $G_{15}$, $G_{30}$ and $G_{45}$, where each gating scheme $G_{d}$ outputs an image of a slice at distance $d$ (see Fig. \ref{fig:gated_principle}). 
By repeating the gating schemes periodically, an image sequence for each distance is obtained at the expense of frame rate divided by three.
Given the duration of the laser pulse and integration time on the chip, the image cannot be captured at a perfect distance. 
Therefore, each slice is additionally characterized by a certain width $w$. 
As the cycle time for each slice is in the microseconds domain, dynamic scenarios will differ only very slightly between the slices. 

Gated imaging systems can be theoretically driven in any wavelength. 
In the \ac{NIR} range, sensors can be relatively easily fabricated using a standard silicon \ac{CMOS} technique and the lasers can be driven by a simple laser diode. 
The main constraint of gated imaging in \ac{NIR} is the maximum allowed laser power due to eye safety regulations. 
Especially in bad weather, high laser power is required in order to illuminate the scene as best as possible. 
In this work, we use a \ac{NIR} gated imaging system from BrightwayVision\footnote{brightwayvision.com} with a gated \ac{CMOS} sensor with resolution 1280x960 pixels. One slice is released with a frame rate of \unit[30]{Hz}.
As the length of the fog chamber is limited, only one slice with parameters given in Table \ref{tab:gating_parameters} is recorded. 
This slice starts from $d = \frac{c \cdot t_\text{delay}}{2} = \unit[13.5]{m}$.

\begin{table}
	\centering
	\caption{Gating parameters.}
	\begin{tabular}{ll} \hline
		parameter & value \\ \hline
		laser on $t_\text{laser}$ & \unit[160]{ns} \\ \hline
		delay $t_\text{delay}$ & \unit[90]{ns} \\ \hline
		exposure $t_\text{gate}$ & \unit[160]{ns} \\ \hline
		micro exposures $m$ & 2000 \\ \hline
	\end{tabular}
	\label{tab:gating_parameters}
\end{table}

%% file: fogchamber_experiment.tex
\section{Fog Chamber Experiment} \label{sec:fog_chamber}

The fog chamber in Clermont Ferrand, operated by CEREMA, provides testing in 
well-defined weather conditions \cite{Colomb2008}.
The chamber is \unit[30]{m} in length, \unit[5.5]{m} wide and \unit[2]{m} high. 
It offers two different fog droplet distributions, i.e. \emph{radiation fog} (small droplets) and \emph{advection fog} (large droplets) with the ability to continuously change the visibility $V$. 
Radiation fog is characterized by a mean diameter of \unit[2]{\textmu m} whereas advection fog has a mean diameter of \unit[6]{\textmu m}. 
The exact droplet distributions are given in \cite{Colomb2008}.

\begin{figure*}
 \centering 
 
\begin{tabular}{>{\centering\arraybackslash}m{0.5cm} 
>{\centering\arraybackslash}m{0.45\textwidth} >{\centering\arraybackslash}m{ 
0.45\textwidth}}
   & 
scenario 1 & 
scenario 2\\

\rotatebox[origin=l]{90}{setup} & 
\includegraphics[width=0.45\textwidth]{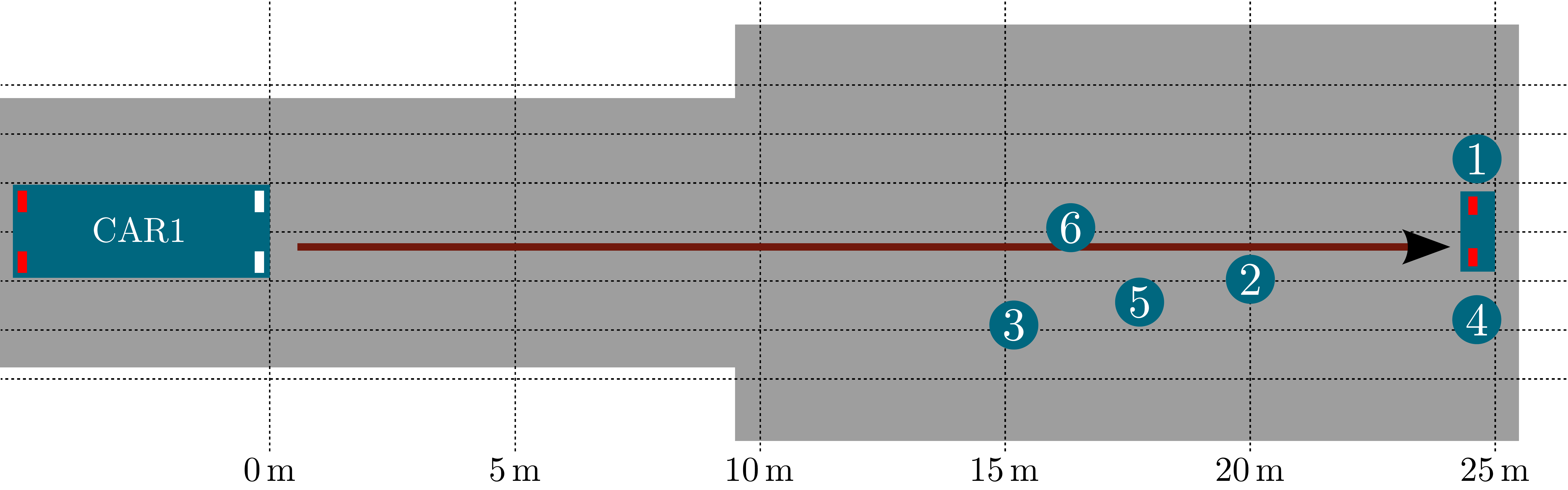} 
& 
\includegraphics[width=0.45\textwidth]{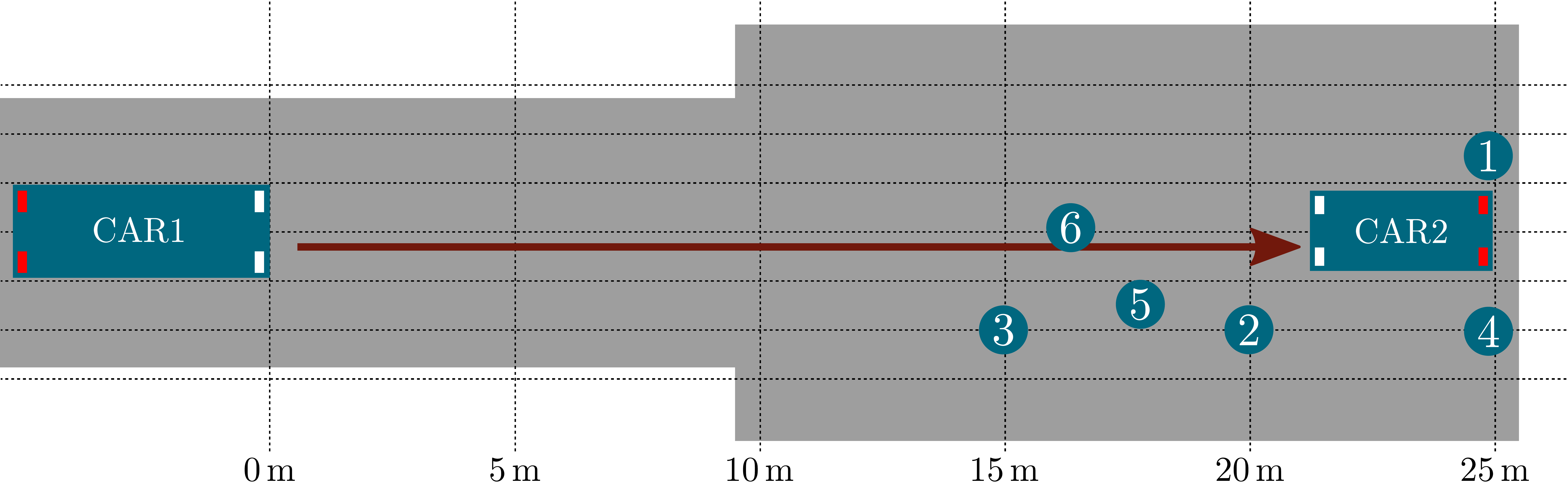} \\

 \rotatebox[origin=l]{90}{reference} & 
 \includegraphics[width=0.45\textwidth]{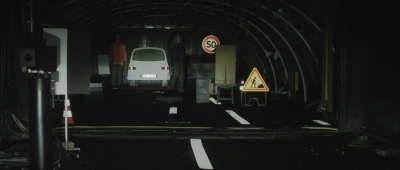} 
 & 
 \includegraphics[width=0.45\textwidth]{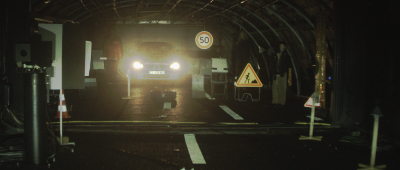} \\
 
   \rotatebox[origin=l]{90}{standard} & 
\includegraphics[width=0.45\textwidth]{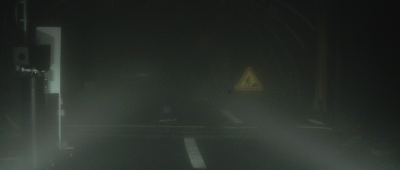} 
& 
\includegraphics[width=0.45\textwidth]{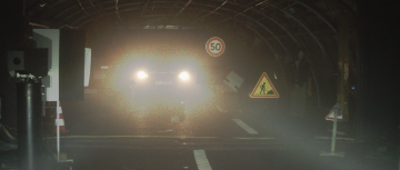} \\

   \rotatebox[origin=l]{90}{gated}  
& 
\includegraphics[width=0.45\textwidth]{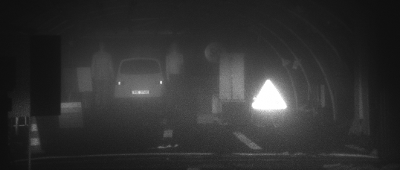} &
\includegraphics[width=0.45\textwidth]{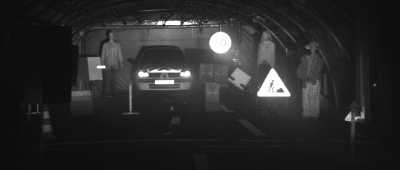} \\

 \end{tabular}
 \caption{Scene setup, reference image of a standard camera without fog and images in fog of a standard camera and a gated camera taken from both scenarios. Scenario 1 at visibility $V\approx\unit[30-40]{m}$. Scenario 2 at visibility $V\approx\unit[50-60]{m}$. The scene setup contains three differently dressed pedestrian mannequins, i.e. man (1), woman (2), child (3), speed limit traffic sign (4) and construction traffic sign (5), a tire (6), road marking and a car (CAR2).}
 \label{fig:scenarios}
\end{figure*}

Many different targets are placed in the fog chamber for qualitative as well as quantitative evaluation, see Fig. \ref{fig:scenarios}. 
Two different test scenarios are considered.
While scenario 1 only deals with passive targets, scenario 2 contains an oncoming car with activated high beams as an active light source that causes extreme scattering.
Halogen lights were selected for the oncoming car because their light distribution has the greatest overlap with the \ac{NIR} region.
For qualitative evaluation, three differently dressed pedestrian mannequins, i.e. man (1), woman (2), child (3), speed limit traffic sign (4) and construction traffic sign (5), a tire (6), road marking and a car (CAR2) are set up. 
For quantitative evaluation, measurements rely on three well-calibrated Zenith 
Polymer diffusive reflectance targets \disable{\footnote{sphereoptics.de}} with a 
reflectivity of 5\,\%, 50\,\% and 90\,\%. 
The Zenith Polymer targets offer a very stable reflectivity from \unit[300]{nm} up to \unit[2500]{nm}. Therefore, different sensors within the visible, \ac{NIR} and even in the \ac{SWIR} region can be easily compared. 

All measurements were performed with activated high beams at CAR1 for the standard camera 
and activated laser illumination for the gated camera. 
We investigated both fog types in four different visibility ranges, \unit[10-20]{m}, \unit[20-30]{m}, \unit[30-40]{m} and \unit[50-60]{m}. 
The fog chamber provides live measurements for the visibility $V$.

In order to obtain a quantitative benchmark between stereo camera and gated camera, we move the reflectance targets at a height of \unit[1.6]{m} along the main viewing axis (red arrow in Fig. \ref{fig:scenarios}) starting from the test vehicle (CAR1) up to the end of the chamber. 
The lidar is used to track the target distance. 
During post processing, a semi-automatic tracking algorithm in combination with human correction delivers the raw intensities on the chip for each reflectance target.
The distance is binned in depth slices and for each depth slice the mean  intensity for each target and its variance is calculated. 

%% file: evaluation.tex
\section{Evaluation}

This section presents different levels of benchmarking. We start with a simple qualitative description of the behavior of standard and gated imaging in fog. Then, we quantize the degeneration by introducing information content as a metric. Finally, we apply multiple evaluations on the mean reflectivity for each target with respect to its depth $d$ for benchmarking scenarios and sensors. 

\input{qualitative_evaluation}
\input{information_content}
\input{reflectance_target}

%% file: qualitative_evaluation.tex
\subsection{Qualitative Evaluation}
\subsubsection{Scenario 1}

Fig.~\ref{fig:scenarios} shows that a passive sensor such as the standard camera suffer heavily from fog with visibility $V\approx\unit[30-40]{m}$ and shows strong contrast degeneration due to attenuation and air-light. 
While in the standard camera only retro reflectors as the construction sign are visible, the gated camera is able to see even the pedestrians at the end of the chamber. 
Due to the concept of gating no backscatter from the car's headlights disturbs the imager.

\subsubsection{Scenario 2}

Scenario 2 with active light-radiating targets is even more challenging. 
Although in Fig. \ref{fig:scenarios} the images are taken from less foggy conditions with visibility $V\approx\unit[50-60]{m}$, it is impossible to perceive objects in front of and close to the car. 
The headlight illumination of CAR2 scattered by the fog creates large light 
blobs which are called corona.
This means a significant risk for vulnerable road users such as pedestrians and cyclists.
Only an active and time-synchronized active sensor such as the gated 
camera is not influenced by an oncoming car. 
The gated camera contains a very narrow bandpass filter that filters out every wavelength except that of the laser illumination.
Moreover, the very short exposure times and the high power laser pulses compared to the air-light help to avoid almost every disturbance of the oncoming car. 

%% file: information_content.tex
\subsection{Information Content Evaluation}

In order to quantify the advantages of gated imaging in foggy conditions, we consider the information content of an image as in \cite{Peynot2009} where they described the information content as a measure to determine when there is not enough information in the sensor stream for multi-sensor fusion. 
The information content $H\left( X \right)$ is a measure in information theory, also known as \emph{entropy}, and is given by \cite{Shannon1948}
\begin{align}
H\left( X \right) = \sum\limits_{i=1}^{n} P\left(x_i\right) \log_2 
P\left(x_i\right)
\end{align}
where $P\left( X \right)$ is the probability mass function of a discrete random 
variable $X$ with possible values $\left\{ x_1, \dots, x_n\right\}$.

The entropy $H$ of an image is obtained by a histogram of all pixel values $\left\{ x_1, \dots, x_n\right\}$. 
Low entropy images usually contain many pixels with the same or similar pixel values and thus have little contrast. 
Images with high contrast and structure contain much entropy. 
It was shown in \cite{Hasirlioglu2016} that rain influence lowers the contrast and brightens the image. 
The entropy is lower-bounded by 0, e.g. completely black image and upper-bounded by the bit depth, i.e. the binary logarithm of the number of possible bit values.

\begin{figure}
	\begin{tikzpicture}
	\begin{axis}[
	xlabel=Visibility $V$ / \unit{m},
	ylabel=Entropy $H\left( X \right)$ /  \unit{bit},
	grid=major,
	legend style={
		cells={anchor=west},
		legend pos=north west,
		font=\footnotesize
	},
	legend entries={standard camera, gated camera},
	width=\columnwidth,
	height=0.65\columnwidth
	]
	
	\addlegendimage{very thick, solid, dai_ligth_grey40K, mark=*}
	\addlegendimage{very thick, solid, dai_petrol, mark=triangle*}
	
	\addplot+ [very thick, solid, dai_ligth_grey40K,mark=*,
	mark size={2},
	mark options={dai_ligth_grey40K}, 
	error bars/.cd,
	y dir=both,
	y explicit, 
	error mark options={rotate=90,dai_ligth_grey40K}
	]
	table[col sep=comma, x={x}, y={y}, y error = {std_y}]{data/entropy_standard.txt};
	
	\addplot+ [very thick, solid, color=dai_petrol,
	mark=triangle*,
	mark size={2}, 
	mark options={dai_petrol},  
	error bars/.cd,
	y dir=both,
	y explicit, 
	error mark options={rotate=90,dai_petrol}
	]
	table[col sep=comma, x={x}, y={y}, y error = {std_y}]{data/entropy_bwv.txt};
	
	\end{axis}
	\end{tikzpicture}
	\caption{Entropy $H\left(X\right)$ vs. visibility generated by 
		three recordings at a certain visibility range (\unit[20-30]{m}, 
		\unit[30-40]{m} and \unit[50-60]{m}). The error bars show the standard deviation 
		of the entropy averaged over a series of images.}
	\label{fig:entropy}
\end{figure}
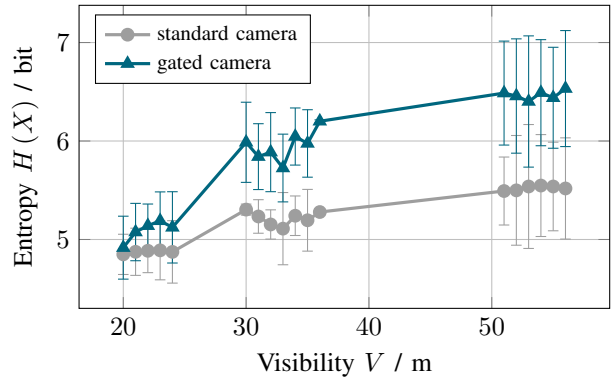

In order to obtain a fair comparison, only the part of the image which 
is illuminated by both headlights and laser is considered. 
Moreover, resolution and pixel depth are adjusted to equality. 
For color images, the entropy can be obtained by calculating the entropy for each color channel and then averaging.
Fig.~\ref{fig:entropy} visualizes the entropy of the standard camera and the 
gated camera subject to the different visibilities for scenario 1. 
It clearly shows that the entropy $H\left(X\right)$ increases with higher 
visibility and that the gated camera provides more information than the 
standard camera.

%% file: reflectance_target.tex
\subsection{Reflectance Target Evaluation}

The following evaluations are based on the captured sensor images. 
The reflectance targets are extracted from the sensor images. 
The mean intensities of the reflectance targets allows to benchmark the sensor itself and compare it to other sensors.
First, we validate our measurements by fitting the measurements into a physical model. 
Then, we compare the intensity values for different types of fog and different visibility ranges. 
Introducing contrast measures clearly shows contrast degeneration due to fog. 
Finally, we show how the mean intensities $I$ with respect to depth $d$ gives deep insight into the performance of image sensors.

\subsubsection{Model fit}

\begin{figure}[t]
	\begin{tikzpicture}
	\begin{axis}[
	xlabel=Depth $d$ / \unit{m},
	ylabel=Intensity $I$ / \%,
	xmin=5,
	xmax=23,
	ymin=0,
	ymax=0.65,    
	grid=major,
	legend entries={5\,\% Target, 50\,\% Target, 90\,\% Target, model fit},
	legend style={
		cells={anchor=west},
		legend pos=north east,
		font=\footnotesize
	},
	width=\columnwidth,
	height=0.65\columnwidth
	]
	
	\addlegendimage{very thick, dai_ligth_grey40K, mark=*}
	\addlegendimage{very thick, dai_deepred, mark=triangle*}
	\addlegendimage{very thick, dai_petrol, mark=square*}
	\addlegendimage{very thick, densely dashed}

	\addplot+ [very thick, solid, dai_ligth_grey40K,mark=*,
	mark size={2},
	mark options={solid,dai_ligth_grey40K}, 
	mark repeat={10},
	error bars/.cd,
	y dir=both,
	y explicit, 
	error mark options={rotate=90,dai_ligth_grey40K}
	]
	table[col sep=comma, x={x_5}, y={mean_5}, y error = {std_5}]{data/rawDataAptinaTest19Day2LightFog50m.txt};
	
	\addplot+ [very thick, solid, color=dai_deepred,
	mark=triangle*,
	mark size={2}, 
	mark options={solid,dai_deepred}, 
	mark repeat={10},
	error bars/.cd,
	y dir=both,
	y explicit, 
	error mark options={rotate=90,dai_deepred}
	]
	table[col sep=comma, x={x_50}, y={mean_50}, y error = {std_50}]{data/rawDataAptinaTest19Day2LightFog50m.txt};
	
	\addplot+ [very thick, solid, color=dai_petrol,
	mark=square*,
	mark size={2}, 
	mark options={solid,dai_petrol},  
	mark repeat={10},
	error bars/.cd,
	y dir=both,
	y explicit, 
	error mark options={rotate=90,dai_petrol}
	]
	table[col sep=comma, x={x_90}, y={mean_90}, y error = {std_90}]{data/rawDataAptinaTest19Day2LightFog50m.txt};
	
	\addplot+ [color=dai_petrol, densely dashed, no markers, very thick
	]
	table[col sep=comma, x={x_5}, y={fit90}]{data/fit.txt};
	
	\addplot+ [color=dai_deepred, densely dashed, no markers, very thick
	]
	table[col sep=comma, x={x_5}, y={fit50}]{data/fit.txt};
	
	\addplot+ [color=dai_ligth_grey40K, densely dashed, no markers, very thick
	]
	table[col sep=comma, x={x_5}, y={fit5}]{data/fit.txt};
	
	\end{axis}
	\end{tikzpicture}
	\caption{Intensity $I$ on the chip of the standard camera with respect to depth 
		$d$ in a foggy scenario with visibility $V \approx \unit[50-60]{m}$ (scenario 1). 
		The intensity model from Eq. \ref{eq:fit_model} is fitted into the measurement.}
	\label{fig:Aptina50Fog}
\end{figure}
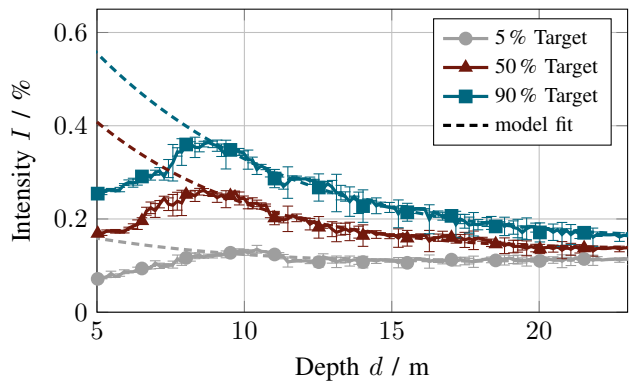

Suppose that the car headlights as well as the laser diodes illuminate the scene in front of the car equally, then the scene radiance can be neglected, i.e. $J(d) \approx I_0$. Moreover, we have to extend the derived model in Eq.~\ref{eq:scattering_model} by an additional shift $d_0$ for the estimated delay until the reflectance targets are fully illuminated by the headlamps and with a different attenuation coefficient $\beta_a$ for the air-light because the targets are mostly not completely within the headlight cone and therefore not fully illuminated.
This gives the adapted scattering model
\begin{align}
I(d)=&I_0 e^{-\beta(d-d_0)} +I_\infty\left(1-e^{-\beta_a(d-d_0)}\right) 
\label{eq:fit_model}
\end{align}
with open parameters $I_0$, $I_\infty$, $d_0$ and $\beta_a$. 
The attenuation $\beta$ can be derived from the meteorological viewing distance $V$ according to Eq.~\ref{eq:visibility} by
\begin{align}
\beta = \frac{log(0.05)}{V} \approx \frac{3}{V}.
\end{align}
In Fig.~\ref{fig:Aptina50Fog}, the fitted model fits the measurements very well for the range where the headlights illuminate the targets.
Additionally, this model allows to simulate greater depths than inside the limited fog chamber by tweaking the parameters sensor resolution, reflectivity ($I_0$), light intensity ($I_{\infty}$) and attenuation coefficient ($\beta$). 

\subsubsection{Reflectivity vs. visibility}
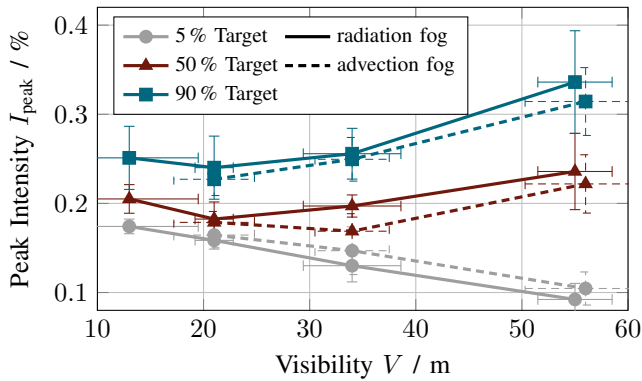
\begin{figure}[t]
  \begin{tikzpicture}
   \begin{axis}[
    xlabel=Visibility $V$ / \unit{m},
    ylabel=Peak Intensity $I_\text{peak}$  / \%,
    grid=major,
    xmin=10,
    xmax=60,
    ymin=0.08,
    ymax=0.42,    
    legend style={
      cells={anchor=west},
      legend pos=north west,
      font=\footnotesize
    },
    legend entries={5\,\% Target, radiation fog, 50\,\% Target, advection fog, 
90\,\% 
Target},
    legend columns=2,
    width=\columnwidth,
    height=0.65\columnwidth
]

\addlegendimage{very thick, dai_ligth_grey40K, mark=*}
\addlegendimage{very thick, solid}
\addlegendimage{very thick, dai_deepred, mark=triangle*}
\addlegendimage{very thick, densely dashed}
\addlegendimage{very thick, dai_petrol, mark=square*}

\addplot+ [very thick,
	   solid,
	   color=dai_ligth_grey40K,
	   mark=*,
           mark size={2},
           mark options={solid, dai_ligth_grey40K}, 
           error bars/.cd,
	   	   y dir = both, y explicit, 
	   x dir = both, x explicit,  
           error mark options={rotate=90,dai_ligth_grey40K}
]
table[col sep=comma, x={dist}, y={Ref_5}, x error = {err}, y error={Ref_5_std}]{data/SmallFogStatsAptina.txt};
   
\addplot+ [very thick,
	   solid,
	   color=dai_deepred,
	   mark=triangle*,
	   mark size={2}, 
	   mark options={solid,dai_deepred},  
	   error bars/.cd,
	   	   y dir = both, y explicit, 
	   x dir = both, x explicit,  
           error mark options={rotate=90,dai_deepred}
]
table[col sep=comma, x={dist}, y={Ref_50}, x error = {err}, y error={Ref_50_std}]{data/SmallFogStatsAptina.txt};

\addplot+ [very thick,
	   solid,
	   color=dai_petrol,
	   mark=square*,
	   mark size={2}, 
	   mark options={solid,dai_petrol},  
	   error bars/.cd,
	   	   y dir = both, y explicit, 
	   x dir = both, x explicit,   
           error mark options={rotate=90,dai_petrol}
]
table[col sep=comma, x={dist}, y={Ref_90}, x error = {err}, y error={Ref_90_std}]{data/SmallFogStatsAptina.txt};

\addplot+ [very thick,
	   densely dashed,
	   color=dai_ligth_grey40K,
	   mark=*,
           mark size={2},
           mark options={solid,dai_ligth_grey40K}, 
           error bars/.cd,
	   	   y dir = both, y explicit, 
	   x dir = both, x explicit, 
           error mark options={rotate=90,dai_ligth_grey40K}
]
table[col sep=comma, x={dist}, y={Ref_5}, x error = {err}, y error={Ref_5_std}]{data/LargeFogStatsAptina.txt};
   
\addplot+ [very thick,
	   densely dashed,
	   color=dai_deepred,
	   mark=triangle*,
	   mark size={2}, 
	   mark options={solid,dai_deepred},  
	   error bars/.cd,
	   	   y dir = both, y explicit, 
	   x dir = both, x explicit, 
       error mark options={rotate=90,dai_deepred}
]
table[col sep=comma, x={dist}, y={Ref_50}, x error = {err}, y error={Ref_50_std}]{data/LargeFogStatsAptina.txt};

\addplot+ [very thick,
	   densely dashed,
	   color=dai_petrol,
	   mark=square*,
	   mark size={2}, 
	   mark options={solid,dai_petrol},  
	   error bars/.cd,
	   	   y dir = both, y explicit, 
	   x dir = both, x explicit,  
           error mark options={rotate=90,dai_petrol}
]
table[col sep=comma, x={dist}, y={Ref_90}, x error = {err}, y error={Ref_90_std}]{data/LargeFogStatsAptina.txt};

  \end{axis}
 \end{tikzpicture}
 \caption{Peak intensity $I_\text{peak}$ on the chip of a standard camera for each reflectance target. 
}
\label{fig:IntensitiesVsVisibility}
\label{fig:IlluminationAptina}
\end{figure}

Fig.~\ref{fig:IlluminationAptina} gives an overview about the behavior of raw reflectance values for different visibilities in two different fog droplet distributions, i.e. advection and radiation fog. 
For each visibility range, type of fog and each reflectance target, the peak intensity $I_\text{peak}$ is determined.
It can be clearly seen that due to attenuation the 90\,\% target appears darker in denser fog. 
Conversely, the 5\,\% target appears brighter in denser fog due to more air-light. 
The 50\,\% target appears approximately the same. 
In total, this degenerates the contrast. 
Within the error bars, this effect is similar for both fog droplet distributions.

\subsubsection{Contrast}
\begin{figure}[t]
  \begin{tikzpicture}
   \begin{axis}[
    xlabel=Visibility $V$ / \unit{m},
    ylabel=Contrast $c$ ,
    grid=major,
    xmin=10,
    xmax=60,
    ymin=0,
    ymax=0.75,    
    legend style={
      cells={anchor=west},
      legend pos=north west,
      font=\footnotesize
    },
    legend entries={Michelson contrast, RMS contrast, radiation fog,  advection 
fog},
    width=\columnwidth,
    height=0.65\columnwidth
]

\addlegendimage{very thick, dai_ligth_grey40K, mark=*}
\addlegendimage{very thick, dai_petrol, mark=triangle*}
\addlegendimage{very thick, solid}
\addlegendimage{very thick, densely dashed}

\addplot+ [very thick,
	   densely dashed,
	   color=dai_ligth_grey40K,
	   mark=*,
       mark size={2},
       mark options={solid,dai_ligth_grey40K}, 
       error bars/.cd,
	   y dir = both, y explicit, 
	   x dir = both, x explicit, 
       error mark options={rotate=90,dai_ligth_grey40K}
]
table[col sep=comma, x={dist_bg}, y={bg_fog_M}, x error = {err_bg}, y error = {err_bg_fog_M}]{data/ContrastAptina_bg.txt};

\addplot+ [very thick,
	   densely dashed,
	   color=dai_petrol,
	   mark=triangle*,
       mark size={2},
       mark options={solid,dai_petrol}, 
       error bars/.cd,
	   y dir = both, 
	   y explicit, 
	   x dir = both, 
	   x explicit, 
       error mark options={rotate=90,dai_petrol}
]
table[col sep=comma, x={dist_bg}, y={bg_fog_R}, x error = {err_bg}, y error = {err_bg_fog_R}]{data/ContrastAptina_bg.txt};

\addplot+ [very thick,
	   solid,
	   color=dai_ligth_grey40K,
	   mark=*,
       mark size={2},
       mark options={solid,dai_ligth_grey40K}, 
       error bars/.cd,
	   y dir = both, y explicit, 
	   x dir = both, x explicit, 
       error mark options={rotate=90,dai_ligth_grey40K}
]
table[col sep=comma, x={dist_sm}, y={sm_fog_M}, x error = {err_sm}, y error = {err_sm_fog_M}]{data/ContrastAptina_sm.txt};

\addplot+ [very thick,
	   solid,
	   color=dai_petrol,
	   mark=triangle*,
           mark size={2},
           mark options={solid,dai_petrol}, 
           error bars/.cd,
	   y dir = both, y explicit, 
	   x dir = both, x explicit, 
           error mark options={rotate=90,dai_petrol}
]
table[col sep=comma, x={dist_sm}, y={sm_fog_R}, x error = {err_sm}, y error = {err_sm_fog_R}]{data/ContrastAptina_sm.txt};

  \end{axis}
 \end{tikzpicture}
 \caption{Michelson contrast given by Eq. \ref{eq:michelson_contrast} and 
\ac{RMS} contrast given by Eq. \ref{eq:rms_contrast} for different visibilities 
$V$ and for the standard camera.}
\label{fig:ContrastAptina}
\end{figure}
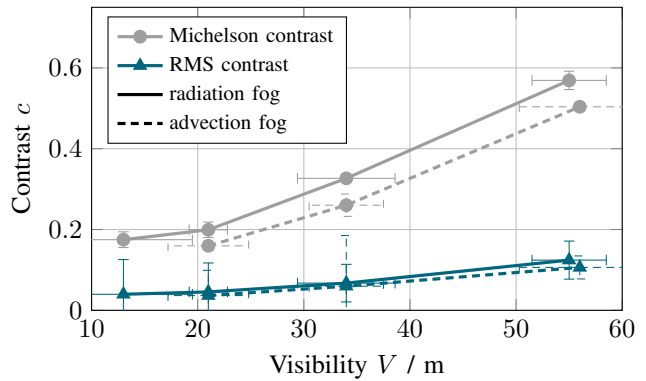

In order to quantize this effect of contrast degeneration, we introduce two 
different contrast measures, the \emph{Michelson contrast} 
\cite{PelliContrastDefinition} given by
\begin{align}
c_{\text{Michelson}} = \frac{I_{90}-I_{5}}{I_{90}+I_{5}}, 
\label{eq:michelson_contrast}
\end{align} 
and the \emph{\ac{RMS} contrast} \cite{PelliContrastDefinition} given by  
\begin{align}
c_{\text{RMS}} = 
\sqrt{\frac{\left(I_{90}-I_{50}\right)^2}{2}+\frac{\left(I_{5}-I_{50}\right)^2}{
2}}. \label{eq:rms_contrast}
\end{align}
The average intensities $I_5$, $I_{50}$ and $I_{90}$ are the measured 
reflectance target 
values taken in between \unit[5-10]{m} in Fig.~\ref{fig:IntensitiesVsVisibility}. 
Fig. \ref{fig:ContrastAptina} illustrates both contrast measures for both fog 
types with respect to visibility. Clearly, increasing the fog 
density decreases the contrast considerable. Textures become indistinguishable and in this way object detection can only be performed by distinguishing the object from the background.

\subsubsection{Comparing Scenarios}

In Fig.~\ref{fig:results_oncomming_Car_Aptina} and \ref{fig:results_oncomming_Car_BWV} the reflectance target measurements for scenario 2 in fog with $V \approx \unit[55]{m}$ are shown. 
Comparing scenario 1 (Fig.~\ref{fig:Aptina50Fog}) with scenario 2 (Fig.~\ref{fig:results_oncomming_Car_Aptina}) as perceived by the standard camera, it can be observed that the standard camera suffers heavily. 
As soon as the targets enter the air-light area of the oncoming car (in Fig.~\ref{fig:results_oncomming_Car_Aptina} starting from $d=\unit[15]{m}$) the contrast between the different targets vanishes very quickly.
Objects can only be detected by differing the object from the background. 
Moreover, the intensity level on the chip increases for scenario 2 with 
the oncoming car compared to scenario 1 without it. 
This can be explained by the scattering that increases the whole scene brightness. 
The intensity peak in Fig.~\ref{fig:results_oncomming_Car_Aptina} at $d=\unit[7]{m}$ gives the same distance as in Fig.~\ref{fig:Aptina50Fog} when the reflectance targets enter the headlight beams.

\subsubsection{Comparing Sensors} 

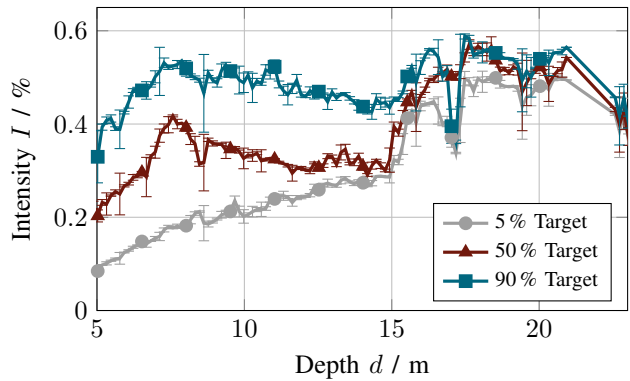
\begin{figure}[t]
\begin{tikzpicture}
   \begin{axis}[
    xlabel=Depth $d$ / \unit{m},
    ylabel=Intensity $I$ / \%,
    xmin=5,
    xmax=23,
    ymin=0,
    ymax=0.65,    
    grid=major,
    legend entries={5\,\% Target, 50\,\% Target, 90\,\% Target},
    legend style={
      cells={anchor=west},
      legend pos=south east,
      font=\footnotesize
    },
    width=\columnwidth,
    height=0.65\columnwidth
]

\addlegendimage{very thick, dai_ligth_grey40K, mark=*}
\addlegendimage{very thick, dai_deepred, mark=triangle*}
\addlegendimage{very thick, dai_petrol, mark=square*}

\addplot+ [very thick, solid, dai_ligth_grey40K,mark=*,
           mark size={2},
           mark options={dai_ligth_grey40K}, 
           mark repeat={10},
           error bars/.cd,
           y dir=both,
           y explicit, 
           error mark options={rotate=90,dai_ligth_grey40K},
]
table[col sep=comma, x={x_5}, y={mean_5}, y error = {std_5}]{data/rawDataAptinaOncommingCarEvaluation50mTest22Day2.txt};
   
\addplot+ [very thick, solid, color=dai_deepred,
	   mark=triangle*,
	   mark size={2}, 
	   mark options={dai_deepred}, 
	   mark repeat={10},
	   error bars/.cd,
	   y dir=both,
	   y explicit, 
	   error mark options={rotate=90,dai_deepred}
]
table[col sep=comma, x={x_50}, y={mean_50}, y error = {std_50}]{data/rawDataAptinaOncommingCarEvaluation50mTest22Day2.txt};

\addplot+ [very thick, solid, color=dai_petrol,
	   mark=square*,
	   mark size={2}, 
	   mark options={dai_petrol}, 
	   mark repeat={10},
	   error bars/.cd,
	   y dir=both,
	   y explicit, 
	   error mark options={rotate=90,dai_petrol}
]
table[col sep=comma, x={x_90}, y={mean_90}, y error = {std_90}]{data/rawDataAptinaOncommingCarEvaluation50mTest22Day2.txt};

  \end{axis}
 \end{tikzpicture}
 \caption{Intensity $I$ on the chip of the standard camera vs. depth $d$ for 
foggy conditions with visibility $V \approx \unit[50-60]{m}$ and an oncoming car 
(scenario 2).}
 \label{fig:results_oncomming_Car_Aptina}
\end{figure}

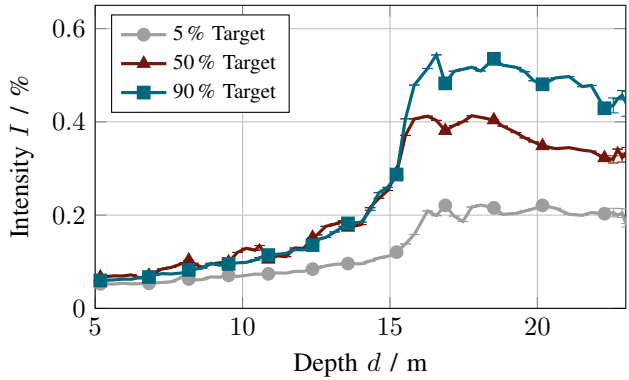
\begin{figure}[t]
  \begin{tikzpicture}
   \begin{axis}[
    xlabel=Depth $d$ / \unit{m},
    ylabel=Intensity $I$ / \%,
    grid=major,
    legend style={
      cells={anchor=west},
      legend pos=north west,
      font=\footnotesize
    },
    legend entries={5\,\% Target, 50\,\% Target, 90\,\% Target},
    xmin=5,
    xmax=23,
    ymin=0,
    ymax=0.65,
    width=\columnwidth,
    height=0.65\columnwidth
]

\addlegendimage{very thick, dai_ligth_grey40K, mark=*}
\addlegendimage{very thick, dai_deepred, mark=triangle*}
\addlegendimage{very thick, dai_petrol, mark=square*}

\addplot+ [very thick, solid, dai_ligth_grey40K,mark=*,
           mark size={2},
           mark options={dai_ligth_grey40K}, 
           mark repeat={5},
           error bars/.cd,
           y dir=both,
           y explicit, 
           error mark options={rotate=90,dai_ligth_grey40K}
]
table[col sep=comma, x={x_5}, y={mean_5}, y error = {std_5}]{data/rawDataBrightwayOncommingCarEvaluation50mTest22Day2.txt};

\addplot+ [very thick, solid, color=dai_deepred,
	   mark=triangle*,
	   mark size={2}, 
	   mark options={dai_deepred},
	   mark repeat={5},
	   error bars/.cd,
	   y dir=both,
	   y explicit, 
	   error mark options={rotate=90,dai_deepred}
]
table[col sep=comma, x={x_50}, y={mean_50}, y error = {std_50}]{data/rawDataBrightwayOncommingCarEvaluation50mTest22Day2.txt};

\addplot+ [very thick, solid, color=dai_petrol,
	   mark=square*,
	   mark size={2}, 
	   mark options={dai_petrol},
	   mark repeat={5},
	   error bars/.cd,
	   y dir=both,
	   y explicit, 
	   error mark options={rotate=90,dai_petrol}
]
table[col sep=comma, x={x_90}, y={mean_90}, y error = {std_90}]{data/rawDataBrightwayOncommingCarEvaluation50mTest22Day2.txt};

  \end{axis}
 \end{tikzpicture}
 \caption{Intensity $I$ on the chip of the gated camera with respect to 
depth $d$ in a foggy scenario with visibility $V \approx \unit[50-60]{m}$ 
(scenario 2).}
 \label{fig:results_oncomming_Car_BWV}
\end{figure}

The gated camera is set up to illuminate an image slice starting from \unit[13.5]{m}.  
Therefore, a steep rise of intensity on the chip is visible in Fig.~\ref{fig:results_oncomming_Car_BWV} at $d=\unit[13.5]{m}$. 
Additionally, a clear difference in intensity on the chip between the reflectance targets can be observed. 
This shows the impressive contrast of gated imaging.
Compared to standard imaging (see Fig.~\ref{fig:results_oncomming_Car_Aptina}), no 
disturbance of the reference targets is visible, although there is an oncoming 
car, thus backscatter is clearly avoided. 
Hence, the measurements of the gated camera for scenario 1 are mostly the same. 
As the length of the fog chamber is limited to \unit[25]{m}, the distance which has to be illuminated is very short and therefore the exposure time and the duration of laser illumination have to be setup short, too. 
Consequently, the lasers have not reached the maximal possible peak power and therefore similar results are expected at greater distances as well.

%% file: outlook_conclusion.tex
\section{Outlook and Conclusion}

This work shows a comprehensive methodology on how to benchmark image sensor behavior in adverse weather such as fog. 
We presented a testing procedure in a fog chamber which allows in addition to qualitative evaluation also quantitative evaluation. 
A physical model was derived which describes the intensity on the chip and enables simulation to further depths as the length of the fog chamber is limited.
For a first evaluation, entropy is a good metric to detect which sensor performs better. 
More elaborated evaluations with calibrated reflectance targets allows to describe the sensor behavior according to the visibility or fog type.
Michelson contrast and \ac{RMS} contrast show the contrast degeneration of a standard camera in fog. 
As we obtain with our methodology raw intensities with respect to depth, it is also possible to compare different sensors to each other. 
This evaluation gives a lot of insight into the precise behavior of each sensor, its advantages and disadvantages. In this way, novel sensor technologies can be tested in adverse weather conditions and compared to state-of-the-art sensors. This methodology also enables to evaluate algorithms which are supposed to improve degraded images.

This evaluation confirms with actual numbers how standard cameras suffer in adverse weather. 
The main disadvantage of standard cameras is the loss of contrast due to air-light and attenuation. This loss of contrast was investigated for two fog droplet distributions which correspond to advection and radiation fog. Within the error bars the loss of contrast is similar for both types. In conclusion, fog degenerates edges within images and strong disturbances appear, which leads to the failure of state-of-the-art algorithms, e.g. for object detection.  
In contrast, the gated camera shows much better contrast and higher viewing distances can be perceived. 
As the illumination distribution within an image slice can be set to be almost constant, the contrast between the illuminated slice and the surrounding is very high. 
This can help to develop more robust and reliable detection algorithms in adverse weather.
Currently, illumination has to be adapted manually to the fog conditions. This may be automated in the future.

The length of the fog chamber limits the evaluation of sensors. 
Future work may investigate how the results in the limited fog chamber can be transformed to larger distances or vice versa, e.g. by physical models. 

%% file: acknowledgment.tex
\section*{Acknowledgment}

The research leading to these results has received funding from the European Union under the H2020 ECSEL Programme as part of the DENSE project, contract number 692449.
We gratefully acknowledge the support and the test facility from our collaboration partners at CEREMA.